\input harvmac
\input graphicx

\def\Title#1#2{\rightline{#1}\ifx\answ\bigans\nopagenumbers\pageno0\vskip1in
\else\pageno1\vskip.8in\fi \centerline{\titlefont #2}\vskip .5in}
%

%%%%%%%%%%%%%%%%%%
%
% Figure macros, SBG 3/03
%
\ifx\includegraphics\UnDeFiNeD\message{(NO graphicx.tex, FIGURES WILL BE IGNORED)}
\def\figin#1{\vskip2in}% blank space instead
\else\message{(FIGURES WILL BE INCLUDED)}\def\figin#1{#1}
\fi
\def\Fig#1{Fig.~\the\figno\xdef#1{Fig.~\the\figno}\global\advance\figno
 by1}
%
%  Ifig   usage:
%
%         \Ifig{\Fig\figlabel}{caption}{figfile}{hsize}
%
% where vsize is the desired vertical size of the figure in truein
%
\def\Ifig#1#2#3#4{
\goodbreak\midinsert
\figin{\centerline{
\includegraphics[width=#4truein]{#3}}}
\narrower\narrower\noindent{\footnotefont
{\bf #1:}  #2\par}
\endinsert
}
%
%defs
%
\font\ticp=cmcsc10

\def\calo{{\cal O}}
\def\calh{{\cal H}}
\def\ehat{{\hat e}}

\def\limrho{\buildrel \rho\rightarrow \pi/2 \over \longrightarrow}

\def\q{{\bf q}}
\def\x{{\bf x}}

\def\roughly#1{\mathrel{\raise.3ex\hbox{$#1$\kern-.75em\lower1ex\hbox{$\sim$}}}}

\font\bbbi=msbm10
\def\bbb#1{\hbox{\bbbi #1}}
%
%refs
%
\lref\Hawkrad{
  S.~W.~Hawking,
  ``Particle Creation By Black Holes,''
  Commun.\ Math.\ Phys.\  {\bf 43}, 199 (1975)
  [Erratum-ibid.\  {\bf 46}, 206 (1976)].
  %%CITATION = CMPHA,43,199;%%
}
\lref\Bousso{
  R.~Bousso, B.~Freivogel, S.~Leichenauer and V.~Rosenhaus,
  ``Eternal inflation predicts that time will end,''
  arXiv:1009.4698 [hep-th].
  %%CITATION = ARXIV:1009.4698;%%
}
\lref\WeinbergNX{
  S.~Weinberg,
  ``Infrared photons and gravitons,''
  Phys.\ Rev.\  {\bf 140}, B516 (1965).
  %%CITATION = PHRVA,140,B516;%%
}
\lref\SGconj{
  Z.~Bern, L.~J.~Dixon and R.~Roiban,
  ``Is N = 8 Supergravity Ultraviolet Finite?,''
  Phys.\ Lett.\  B {\bf 644}, 265 (2007)
  [arXiv:hep-th/0611086].
  %%CITATION = PHLTA,B644,265;%%
}
\lref\EaGi{
  D.~M.~Eardley and S.~B.~Giddings,
  ``Classical black hole production in high-energy collisions,''
  Phys.\ Rev.\  D {\bf 66}, 044011 (2002)
  [arXiv:gr-qc/0201034].
  %%CITATION = PHRVA,D66,044011;%%
}
\lref\MuSo{
  I.~J.~Muzinich and M.~Soldate,
  ``High-Energy Unitarity of Gravitation and Strings,''
  Phys.\ Rev.\  D {\bf 37}, 359 (1988).
  %%CITATION = PHRVA,D37,359;%%
}
\lref\VeVe{
  H.~L.~Verlinde and E.~P.~Verlinde,
  ``Scattering at Planckian energies,''
  Nucl.\ Phys.\  B {\bf 371}, 246 (1992)
  [arXiv:hep-th/9110017].
  %%CITATION = NUPHA,B371,246;%%
}
\lref\KabatTB{
  D.~Kabat and M.~Ortiz,
  ``Eikonal Quantum Gravity And Planckian Scattering,''
  Nucl.\ Phys.\  B {\bf 388}, 570 (1992)
  [arXiv:hep-th/9203082].
  %%CITATION = NUPHA,B388,570;%%
}
\lref\ACV{
  D.~Amati, M.~Ciafaloni, G.~Veneziano,
  ``Classical and Quantum Gravity Effects from Planckian Energy Superstring Collisions,''
Int.\ J.\ Mod.\ Phys.\  {\bf A3}, 1615-1661 (1988).
%%CITATION = CERN-TH-4886/87%%
}
\lref\ACVmore{ D.~Amati, M.~Ciafaloni and G.~Veneziano,
  ``Can Space-Time Be Probed Below The String Size?,''
  Phys.\ Lett.\  B {\bf 216}, 41 (1989)\semi
  %%CITATION = PHLTA,B216,41;%%
  D.~Amati, M.~Ciafaloni and G.~Veneziano,
  ``Higher Order Gravitational Deflection And Soft Bremsstrahlung In Planckian
  Energy Superstring Collisions,''
  Nucl.\ Phys.\  B {\bf 347}, 550 (1990)\semi
  %%CITATION = NUPHA,B347,550;%%
  D.~Amati, M.~Ciafaloni and G.~Veneziano,
  ``Effective action and all order gravitational eikonal at Planckian
  energies,''
  Nucl.\ Phys.\  B {\bf 403}, 707 (1993).
  %%CITATION = NUPHA,B403,707;%%
}
\lref\SGinfo{S.~B.~Giddings,
  ``Quantum mechanics of black holes,''
  arXiv:hep-th/9412138\semi
  %%CITATION = HEP-TH 9412138;%%
  ``The black hole information paradox,''
  arXiv:hep-th/9508151.
  %%CITATION = HEP-TH 9508151;%%
}
\lref\thooholo{
  G.~'t Hooft,
  ``Dimensional reduction in quantum gravity,''
  arXiv:gr-qc/9310026.
  %%CITATION = GR-QC/9310026;%%
}
\lref\sussholo{
  L.~Susskind,
  ``The World As A Hologram,''
  J.\ Math.\ Phys.\  {\bf 36}, 6377 (1995)
  [arXiv:hep-th/9409089].
  %%CITATION = JMAPA,36,6377;%%
}
\lref\Venezesc{
  G.~Veneziano,
  ``String-theoretic unitary S-matrix at the threshold of black-hole production,''
JHEP {\bf 0411}, 001 (2004).
[hep-th/0410166].
%%CITATION = hep-th/0410166%%
}
\lref\GiNLvC{
  S.~B.~Giddings,
  ``Nonlocality versus complementarity: A Conservative approach to the information problem,''
Class.\ Quant.\ Grav.\  {\bf 28}, 025002 (2011).
[arXiv:0911.3395 [hep-th]].
%%CITATION = arXiv:0911.3395%%
}
\lref\BPS{
  T.~Banks, L.~Susskind and M.~E.~Peskin,
  ``Difficulties For The Evolution Of Pure States Into Mixed States,''
  Nucl.\ Phys.\  B {\bf 244}, 125 (1984).
  %%CITATION = NUPHA,B244,125;%%
}
\lref\STU{
  L.~Susskind, L.~Thorlacius and J.~Uglum,
  ``The Stretched Horizon And Black Hole Complementarity,''
  Phys.\ Rev.\  D {\bf 48}, 3743 (1993)
  [arXiv:hep-th/9306069].
  %%CITATION = PHRVA,D48,3743;%%
}
\lref\tHoodom{
  G.~'t Hooft,
  ``Graviton Dominance in Ultrahigh-Energy Scattering,''
Phys.\ Lett.\  {\bf B198}, 61-63 (1987).
}
\lref\wabhip{
  S.~B.~Giddings,
  ``Why Aren't Black Holes Infinitely Produced?,''
  Phys.\ Rev.\  D {\bf 51}, 6860 (1995)
  [arXiv:hep-th/9412159].
  %%CITATION = PHRVA,D51,6860;%%
}
\lref\BaFi{
  T.~Banks, W.~Fischler,
  ``A Model for high-energy scattering in quantum gravity,''
[hep-th/9906038].
%%CITATION = hep-th/9906038%%
}
\lref\Susstrouble{
  L.~Susskind,
  ``Trouble For Remnants,''
  arXiv:hep-th/9501106.
  %%CITATION = HEP-TH/9501106;%%
}
\lref\Astrorev{
  A.~Strominger,
  ``Les Houches lectures on black holes,''
  arXiv:hep-th/9501071.
  %%CITATION = HEP-TH 9501071;%%
}
\lref\GGP{
 M.~Gary, S.~B.~Giddings, J.~Penedones,
  ``Local bulk S-matrix elements and CFT singularities,''
Phys.\ Rev.\  {\bf D80}, 085005 (2009).
[arXiv:0903.4437 [hep-th]].
%%CITATION = arXiv:0903.4437%%
}
\lref\GiSl{
  S.~B.~Giddings, M.~S.~Sloth,
  ``Semiclassical relations and IR effects in de Sitter and slow-roll space-times,''
JCAP {\bf 1101}, 023 (2011).
[arXiv:1005.1056 [hep-th]].
%%CITATION = arXiv:1005.1056%%
}
\lref\GMH{
  S.~B.~Giddings, D.~Marolf and J.~B.~Hartle,
  ``Observables in effective gravity,''
  Phys.\ Rev.\  D {\bf 74}, 064018 (2006)
  [arXiv:hep-th/0512200].
  %%CITATION = PHRVA,D74,064018;%%
}
\lref\GaGi{
  M.~Gary and S.~B.~Giddings,
  ``The flat space S-matrix from the AdS/CFT correspondence?,''
  Phys.\ Rev.\  D {\bf 80}, 046008 (2009)
  [arXiv:0904.3544 [hep-th]].
  %%CITATION = PHRVA,D80,046008;%%
}
\lref\GSA{
  S.~B.~Giddings, M.~Schmidt-Sommerfeld and J.~R.~Andersen,
  ``High energy scattering in gravity and supergravity,''
  Phys.\ Rev.\  D {\bf 82}, 104022 (2010)
  [arXiv:1005.5408 [hep-th]].
  %%CITATION = PHRVA,D82,104022;%%
}
\lref\GaGiobs{
  M.~Gary and S.~B.~Giddings,
  ``Relational observables in 2d quantum gravity,''
  arXiv:hep-th/0612191, Phys.\ Rev.\  D {\bf 75} 104007 (2007).
  %%CITATION = HEP-TH/0612191;%%
}
\lref\GaGitoapp{M. Gary and S.~B.~Giddings, to appear.}
\lref\GiSr{
  S.~B.~Giddings and M.~Srednicki,
  ``High-energy gravitational scattering and black hole resonances,''
  Phys.\ Rev.\  D {\bf 77}, 085025 (2008)
  [arXiv:0711.5012 [hep-th]].
  %%CITATION = PHRVA,D77,085025;%%
}
\lref\GiPo{
 S.~B.~Giddings and R.~A.~Porto,
  ``The gravitational S-matrix,''
  Phys.\ Rev.\  D {\bf 81}, 025002 (2010)
  [arXiv:0908.0004 [hep-th]].
  %%CITATION = ARXIV:0908.0004;%%
}
\lref\STU{
  L.~Susskind, L.~Thorlacius and J.~Uglum,
  ``The Stretched Horizon And Black Hole Complementarity,''
  Phys.\ Rev.\  D {\bf 48}, 3743 (1993)
  [arXiv:hep-th/9306069].
  %%CITATION = PHRVA,D48,3743;%%
}
\lref\LPSTU{
  D.~A.~Lowe, J.~Polchinski, L.~Susskind, L.~Thorlacius and J.~Uglum,
  ``Black hole complementarity versus locality,''
  Phys.\ Rev.\  D {\bf 52}, 6997 (1995)
  [arXiv:hep-th/9506138].
  %%CITATION = PHRVA,D52,6997;%%
}
\lref\PolS{
  J.~Polchinski,
  ``S-matrices from AdS spacetime,''
  arXiv:hep-th/9901076.
  %%CITATION = HEP-TH/9901076;%%
}
\lref\SussS{
  L.~Susskind,
 ``Holography in the flat space limit,''
  arXiv:hep-th/9901079.
  %%CITATION = HEP-TH/9901079;%%
}
\lref\GKP{
  S.~S.~Gubser, I.~R.~Klebanov and A.~M.~Polyakov,
  ``Gauge theory correlators from non-critical string theory,''
  Phys.\ Lett.\  B {\bf 428}, 105 (1998)
  [arXiv:hep-th/9802109].
  %%CITATION = PHLTA,B428,105;%%
}
\lref\WittAdS{
  E.~Witten,
  ``Anti-de Sitter space and holography,''
  Adv.\ Theor.\ Math.\ Phys.\  {\bf 2}, 253 (1998)
  [arXiv:hep-th/9802150].
  %%CITATION = 00203,2,253;%%
}
\lref\GiNe{
  S.~B.~Giddings and W.~M.~Nelson,
  ``Quantum emission from two-dimensional black holes,''
  Phys.\ Rev.\  D {\bf 46}, 2486 (1992)
  [arXiv:hep-th/9204072].
  %%CITATION = PHRVA,D46,2486;%%
}
\lref\Page{
  D.~N.~Page,
  ``Information in black hole radiation,''
  Phys.\ Rev.\ Lett.\  {\bf 71}, 3743 (1993)
  [arXiv:hep-th/9306083].
  %%CITATION = PRLTA,71,3743;%%
}
\lref\QBHB{
  S.~B.~Giddings,
  ``Quantization in black hole backgrounds,''
  Phys.\ Rev.\  D {\bf 76}, 064027 (2007)
  [arXiv:hep-th/0703116].
  %%CITATION = PHRVA,D76,064027;%%
}
\lref\Hawkunc{
  S.~W.~Hawking,
  ``Breakdown Of Predictability In Gravitational Collapse,''
  Phys.\ Rev.\  D {\bf 14}, 2460 (1976).
  %%CITATION = PHRVA,D14,2460;%%
}
\lref\GiddingsNC{
  S.~B.~Giddings and M.~S.~Sloth,
  ``Semiclassical relations and IR effects in de Sitter and slow-roll
  space-times,''
  arXiv:1005.1056 [hep-th].
  %%CITATION = ARXIV:1005.1056;%%
}
\lref\LQGST{
  S.~B.~Giddings,
  ``Locality in quantum gravity and string theory,''
  Phys.\ Rev.\  D {\bf 74}, 106006 (2006)
  [arXiv:hep-th/0604072].
  %%CITATION = PHRVA,D74,106006;%%
}
\lref\Katzetal{
  A.~L.~Fitzpatrick, E.~Katz, D.~Poland and D.~Simmons-Duffin,
  ``Effective Conformal Theory and the Flat-Space Limit of AdS,''
  arXiv:1007.2412 [hep-th].
  %%CITATION = ARXIV:1007.2412;%%
}
\lref\Duff{ M.~J.~Duff,
  ``Covariant gauges and point sources in general relativity,''
  Annals Phys.\  {\bf 79}, 261 (1973)\semi
  %%CITATION = APNYA,79,261;%% 
  ``Quantum tree graphs and the Schwarzschild solution," Phys.\ Rev.\ {\bf D7} 2317 (1973).}
\lref\GGM{
  S.~B.~Giddings, D.~J.~Gross and A.~Maharana,
  ``Gravitational effects in ultrahigh-energy string scattering,''
  Phys.\ Rev.\  D {\bf 77}, 046001 (2008)
  [arXiv:0705.1816 [hep-th]].
  %%CITATION = PHRVA,D77,046001;%%
}
\lref\BSM{
  S.~B.~Giddings,
  ``The boundary S-matrix and the AdS to CFT dictionary,''
  Phys.\ Rev.\ Lett.\  {\bf 83}, 2707 (1999)
  [arXiv:hep-th/9903048].
  %%CITATION = PRLTA,83,2707;%%
}
\lref\FSS{
  S.~B.~Giddings,
  ``Flat-space scattering and bulk locality in the AdS/CFT  correspondence,''
  Phys.\ Rev.\  D {\bf 61}, 106008 (2000)
  [arXiv:hep-th/9907129].
  %%CITATION = PHRVA,D61,106008;%%
}
\lref\HPPS{
  I.~Heemskerk, J.~Penedones, J.~Polchinski and J.~Sully,
  ``Holography from Conformal Field Theory,''
  JHEP {\bf 0910}, 079 (2009)
  [arXiv:0907.0151 [hep-th]].
  %%CITATION = JHEPA,0910,079;%%
}
\lref\AiSe{
  P.~C.~Aichelburg and R.~U.~Sexl,
  ``On the Gravitational field of a massless particle,''
  Gen.\ Rel.\ Grav.\  {\bf 2}, 303 (1971).
  %%CITATION = GRGVA,2,303;%%
}
\lref\NumBH{
  M.~W.~Choptuik and F.~Pretorius,
  ``Ultra Relativistic Particle Collisions,''
  Phys.\ Rev.\ Lett.\  {\bf 104}, 111101 (2010)
  [arXiv:0908.1780 [gr-qc]].
  %%CITATION = PRLTA,104,111101;%%
}
\lref\Venezfriends{
  D.~Amati, M.~Ciafaloni, G.~Veneziano,
  ``Towards an S-matrix description of gravitational collapse,''
JHEP {\bf 0802}, 049 (2008).
[arXiv:0712.1209 [hep-th]].
%%CITATION = arXiv:0712.1209%%
}
\lref\ReSi{
  M.~Reed and B.~Simon,
  ``Methods Of Mathematical Physics. Vol. 3: Scattering Theory,''
  %\href{http://www.slac.stanford.edu/spires/find/hep/www?irn=913014}{SPIRES entry}
  {\it  New York, USA: Academic (1979) 463p}.
}
\lref\CGHS{
  C.~G.~Callan, Jr., S.~B.~Giddings, J.~A.~Harvey {\it et al.},
  ``Evanescent black holes,''
Phys.\ Rev.\  {\bf D45}, 1005-1009 (1992).
[hep-th/9111056].
%%CITATION = hep-th/9111056%%
}
\lref\Dono{
  J.~F.~Donoghue, T.~Torma,
  ``Infrared behavior of graviton-graviton scattering,''
Phys.\ Rev.\  {\bf D60}, 024003 (1999).
[hep-th/9901156].
%%CITATION = hep-th/9901156%%
}
\lref\MAGOO{
  O.~Aharony, S.~S.~Gubser, J.~M.~Maldacena, H.~Ooguri and Y.~Oz,
 ``Large N field theories, string theory and gravity,''
  Phys.\ Rept.\  {\bf 323}, 183 (2000)
  [arXiv:hep-th/9905111].
  %%CITATION = PRPLC,323,183;%%
}
\lref\HoADSCMT{
  G.~T.~Horowitz,
  ``Surprising Connections Between General Relativity and Condensed Matter,''
  arXiv:1010.2784 [gr-qc].
  %%CITATION = ARXIV:1010.2784;%%
}
\lref\MaroUH{
  D.~Marolf,
  ``Unitarity and Holography in Gravitational Physics,''
  Phys.\ Rev.\  D {\bf 79}, 044010 (2009)
  [arXiv:0808.2842 [gr-qc]].
  %%CITATION = PHRVA,D79,044010;%%
}
\lref\Vene{
  G.~Veneziano,
  ``A Stringy Nature Needs Just Two Constants,''
  Europhys.\ Lett.\  {\bf 2}, 199 (1986).
  %%CITATION = EULEE,2,199;%%
}
\lref\Gross{
  D.~J.~Gross,
  ``Superstrings And Unification,'' PUPT-1108
{\it Plenary Session talk given at 24th Int. Conf. on High Energy Physics, Munich, West Germany, Aug 4-10, 1988}.
  %%CITATION = C88-08-04;%%
}
\lref\locbdrefs{
  S.~B.~Giddings and M.~Lippert,
  ``Precursors, black holes, and a locality bound,''
  Phys.\ Rev.\  D {\bf 65}, 024006 (2002)
  [arXiv:hep-th/0103231]\semi
  ``The information paradox and the locality bound,''
  Phys.\ Rev.\  D {\bf 69}, 124019 (2004)
  [arXiv:hep-th/0402073]\semi
  %%CITATION = PHRVA,D69,124019;%%
  %%CITATION = PHRVA,D65,024006;%%
   S.~B.~Giddings, ``Black hole information, unitarity, and nonlocality,''
  Phys.\ Rev.\  D {\bf 74}, 106005 (2006)
  [arXiv:hep-th/0605196]\semi
  %%CITATION = PHRVA,D74,106005;%%
  ``(Non)perturbative gravity, nonlocality, and nice slices,''
  Phys.\ Rev.\  D {\bf 74}, 106009 (2006)
  [arXiv:hep-th/0606146].
%%CITATION = PHRVA,D74,106009;%%
}
\lref\SekinoHE{
  Y.~Sekino, L.~Susskind,
  ``Fast Scramblers,''
JHEP {\bf 0810}, 065 (2008).
[arXiv:0808.2096 [hep-th]].
%%CITATION = arXiv:0808.2096%%
}
\lref\GiMa{
  S.~B.~Giddings, D.~Marolf,
  ``A Global picture of quantum de Sitter space,''
Phys.\ Rev.\  {\bf D76}, 064023 (2007).
[arXiv:0705.1178 [hep-th]].
%%CITATION = arXiv:0705.1178%%
}
\lref\Erice{
  S.~B.~Giddings,
  ``The gravitational S-matrix: Erice lectures,''
[arXiv:1105.2036 [hep-th]].
%%CITATION = arXiv:1105.2036%%
}
\Title{
%\vbox{\hbox{Draft -- do not distribute}}
\vbox{\baselineskip12pt
}}
{\vbox{\centerline{Is string theory a theory of quantum gravity?}
}}
\centerline{{\ticp 
Steven B. Giddings\footnote{$^\ast$}{Email address: giddings@physics.ucsb.edu}  
} }
\centerline{\sl Department of Physics}
\centerline{\sl University of California}
\centerline{\sl Santa Barbara, CA 93106}
\vskip.10in
\centerline{\bf Abstract}
Some problems in finding a complete quantum theory incorporating gravity are discussed.  One  is that of giving a consistent unitary description of high-energy scattering.  Another is that of giving a consistent quantum description of cosmology, with appropriate observables.  While string theory addresses some problems of quantum gravity, its ability to resolve these remains unclear.  Answers may require new mechanisms and constructs, whether within string theory, or in another framework.
 
\vskip.3in
%\draftmode
\Date{}

\newsec{The problem of gravity}

I have purposefully chosen a controversial title for this contribution.  But, the bar is high -- the challenges of quantum gravity are profound.  While string theory has certainly provided new ideas about how to address some of these problems, it seems clear that it still has a fair ways to go if it is to resolve them all.  

Nonrenormalizability has long been regarded as a central problem of quantum gravity, and here string theory yielded promising success.  Specifically, string theory naturally regulates the infinite proliferation of ultraviolet divergences in the loop expansion, and, order-by-order in perturbation theory, apparently gives UV finite scattering amplitudes.

Another central issue for gravity research has been that  of spacetime singularities.  Here, string theory has also had interesting success, providing new mechanisms and suggestions about how stringy effects can replace singularities by regular configurations, particularly in the case of timelike singularities.  (Spacelike singularities have remained more problematic.)

These successes overcame problems that had been the focus of years of struggle.  But, with time, it has also become clear that these ``short-distance" problems are not the only problems of quantum gravity.  With evolution of our knowledge, it now seems that there are other problems both more ``long-distance," and more profound.

One such deeper problem is that of unitarity of the S-matrix for gravitational scattering.  Other historical examples of incomplete theories -- for example the four-Fermi theory of weak interactions -- could be viewed as presenting a problem of nonrenormalizability or of unitarity, with these being linked, as one finds when studying scattering at energies near the electroweak scale.  Attempts to describe gravitational scattering also encounter a unitarity problem, but one that appears to be associated with long-distance physics, and does not have an obvious cure in any short-distance modification of the theory.  One guise of this problem is the ``black hole information problem" (or ``paradox"), which arose from Hawking's discovery of black hole evaporation\refs{\Hawkrad} and arguments that it implies fundamental nonunitarity\refs{\Hawkunc}.  It is apparently a long-distance problem because, for example, ultraplanckian scattering classically produces large black holes, and one finds a conflict between basic principles seemingly independent of the assumptions made about short distance physics.  Either in this context, or in the context of collapse of a massive object, unitarity appears at odds with locality at macroscopic scales --  as large as the black hole in question.

String theorists have suggested a picture for restoration of unitarity -- through dualities, and the idea of ``holography," but a critical question is whether the proposed picture does in fact furnish a complete description of such gravitational scattering, and how it does so.  

In trying to answer this question, one begins to make contact with another profound problem of quantum gravity:  how does a theory of quantum gravity describe observations (or observables) that are local, at least in some approximation?  In fact, a description of such quantities seems critical for fully addressing the information problem, since the basic ``paradox" results from calculations involving local quantities both inside and outside the black hole.  Moreover, a complete physical picture should be able to describe the observations of an observer who falls into a black hole.  However, existing string constructions so far provide us with (at best) the S-matrix, and do not yet address the general question of providing approximately local observables.

More broadly, the problem of local observables in gravity appears a deep one, and one that touches on other difficulties resulting from string theory.  First, while in field theory local fields provide us with local observables, the diffeomorphism symmetry apparent in a low-energy description of gravity indicates that local observables aren't gauge invariant.  A proposed resolution of this --  whose semiclassical form is used in current studies of inflationary cosmology --  is that local observables are approximately recovered {\it relationally}, specifically in relation  to localized features of the state.  But, such a picture encounters challenges in a full quantum treatment, since the corresponding ``reference background" can fluctuate.  This indicates, at the least, that familiar locality is not a precise concept.

More than that, in a complete relational description one might include   a description of observers as part of the quantum system, but in the context of inflationary cosmology quantum fluctuations can apparently produce new observers.  This problem has become particularly manifest in attempts to make sense of eternal inflation in the ``string landscape," where attempts to regulate corresponding infinities and provide a ``measure" have seemingly lead to nonsensical results, such as an end to time\refs{\Bousso}.  It should be stressed that these problems are also long-distance, or infrared, issues.  We might also note  that these problems are particularly forced upon us by the apparently huge number of possible vacua of string theory, which then requires an environmental/anthropic approach to prediction, and hence discussion of such measures.  (One may also consider the possibility that the vacuum problem is another warning that string theory is missing something important.)

String theory has undoubtedly had many successes, from important implications in mathematics, to giving powerful methods to generate approximate effective descriptions of complex systems such as the quark-gluon plasma or condensed-matter systems\HoADSCMT.  It moreover offers us hints about how to generate the chiral gauge structure of the Standard Model of particle physics.  It is clearly a powerful mathematical framework.

But, if it is to be a fundamental theory of nature, string theory also needs to provide a means to answer the most profound questions of quantum gravity, and these are very challenging.  This contribution will give a more in-depth discussion of some of these challenges, and discuss string theory's response to these. I will also comment on some possible general features of a theory of quantum gravity, whether this arises from string theory, or from another approach.

\newsec{The gravitational S-matrix and the ultraplanckian regime}

\subsec{The S-matrix}

In physics one seeks sharply predictable quantities, and a first question is what these might be in a quantum theory of gravity. We will return to the question of local observables, but the most straightforward candidates for sharp prediction are elements of the S-matrix.\foot{For further discussion, see \Erice.}  This is strongly motivated by empirical observation.  First, the curvature radius of the Universe is of order $10^{60}$ in Planck units, so Minkowski space is an extremely good approximate solution to the underlying theory.  Secondly, we observe that the excitations about Minkowski space involve asymptotic states that are well-described as particles -- electrons, photons, {\it etc.}  So, whatever one has as an underlying fundamental description of these particle states, one expects to be able to give amplitudes for scattering of ``in" states consisting of widely-separated particles to similar ``out" states.   

Even here, there are some subtleties, due to the masslessness of the graviton.  Specifically, in spacetime dimension $D=4$ there are divergences associated with emission of arbitrarily soft gravitons.  While these are infrared divergences, there is a case that they don't drive at more profound issues, but are rather of a more technical nature.  In particular, very similar divergences arise in quantum electrodynamics, and are cured there by summing over soft photons, to give inclusive probabilities that take into account in-practice limitations on any experiment's IR resolving power.  Gravity seems to behave similarly\refs{\WeinbergNX,\Dono}.  Moreover, there are indications that gravity might be formulated in higher dimensions, and there soft-graviton divergences are absent.  

Indeed, string theory directly yields approximate calculations of the S-matrix, in higher-dimensional spacetime, that appear to avoid some pitfalls of earlier approaches to the problem of gravitational scattering.  

A particularly important testing ground for any candidate gravity theory's predictions of the S-matrix is  the ultraplanckian regime, where one scatters particles far above the $D$-dimensional Planck mass $M_D$.\foot{We adopt the convention $M_D^{D-2} = (2\pi)^{D-4}/(8\pi G_D)$.}  That such a limit should exist follows from very general principles: Lorentz invariance of Minkowski space, and a very mild form of locality.  A given particle may be boosted to arbitrarily high energies by a Lorentz transform, and one may moreover consider independent boosts of widely-separated particles, to arrange particles with an ultraplanckian center of mass (CM) energy, which then collide.\foot{Even theories without Lorentz invariance, {\it e.g.} with modified dispersion relations, appear to need to address such a regime, or confront even greater complications.  One example would be in black hole formation from non-relativistic particles.}   

To understand how string theory meets the challenges of this regime, we first need to review some basic features of gravitational scattering.  Some early discussions of  scattering and string theory in this regime include \refs{\tHoodom,\ACV,\ACVmore,\BaFi}.

\subsec{Perturbative scattering}

In the subplanckian regime, Einstein's theory, viewed as an effective field theory, apparently makes perfectly good predictions.   While nonrenormalizability tells us that infinitely many higher-dimension operators are induced through quantization, these naturally have coefficients that are inverse powers of $M_D$, and so make negligible contributions to low-energy gravitational scattering.  Specifically, it appears that in this regime, the Born approximation, namely single graviton exchange, gives a very good approximation to the gravitational S-matrix.  For scattering of particles with incoming (outgoing) momenta $p_1, p_2$ ($-p_3,-p_4$), and using the Mandelstam parameters 
\eqn\mand{
s=-(p_1+p_2)^2\ ,\ t=-(p_1+p_3)^2\ ,\ u=-(p_1+p_4)^2\ ,}
the scattering amplitude in the limit $s\gg -t$ is
\eqn\born{T_{\rm 0} = -8\pi G_D s^2/t\ .}

Conversely, when scattering energies reach the Planck energy, one expects the higher dimension operators to make important and unpredictable contributions; at the same time one expects a breakdown of unitarity, due to growth of amplitudes.  String theory suggests a way to tame these amplitudes, by regulating higher-loop divergences, and effectively fixing the coefficients of the higher-dimension operators.

However, it is important to think about this in a more refined view, where one more carefully specifies scattering parameters.  For example, one might consider scattering as a function of both energy $E$, and impact parameter $b$.  Even for ultraplanckian scattering energies, at sufficiently large impact parameter gravity is very weak, and the Born approximation appears good.  Ultimately, though, for a fixed ultraplanckian energy, at small enough impact parameter the Born approximation does break down.  

 \Ifig{\Fig\Scattdiag}{A proposed ``phase diagram" of different regimes for gravitational scattering.  In particular, we consider the effect of decreasing impact parameter, at fixed ultraplanckian energy, as indicated by the arrow.  NR indicates the regime where higher-dimension operators are expected to be important.}{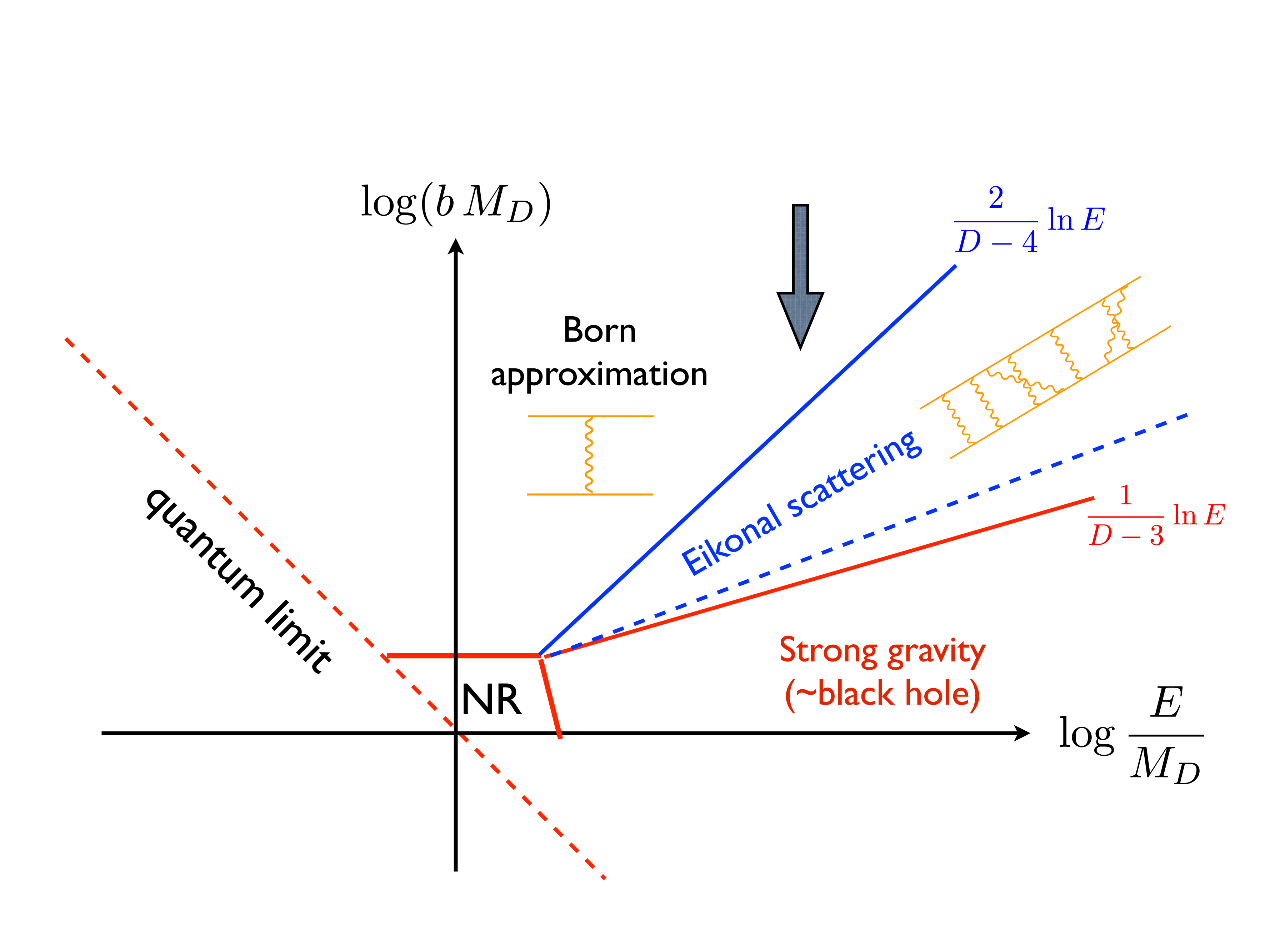}{6.5}

In fact, this happens in the regime of familiar gravity experiments.  Higher-loop corrections become  important,  but these are very specific higher-loop corrections. They correspond to iterated single-graviton exchange, which constructs the ladder (and crossed-ladder) diagrams.  The region where these are important is indicated in \Scattdiag.  When the momentum transfer is small, $q\ll E$, these diagrams can be approximated and summed to give the eikonal amplitudes\refs{\ACV,\ACVmore,\MuSo,\VeVe,\KabatTB},
\eqn\eikamp{iT_{\rm eik}(s,t) = 2s \int d^{D-2} x_\perp e^{-i\q_\perp \cdot \x_\perp}(e^{i\chi(E,x_\perp)} -1)\ ,}
where the eikonal phase $\chi(E,x_\perp)$ is a Fourier transform of the tree-level amplitude,   
\eqn\eikphase{\chi(E, x_\perp) = {1\over 2s}\int{d^{D-2}q_\perp\over(2\pi)^{D-2}}\,
e^{-i{\bf q}_\perp\cdot \x_\perp}T_{\rm tree}(s,-q^2_\perp)}
with respect to the transverse components of the momentum transfer, ${\bf q}_\perp$.
Using the Born (single-graviton) amplitude \born, one easily finds
\eqn\eikexpress{ \chi = {4\pi\over (D-4) \Omega_{D-3}} {G_D s\over x_\perp^{D-4}} }
with $\Omega_n$ the volume of the unit $n$-sphere.
(The exponent $D-4$ instead gives logarithms in $D=4$.)  The classical approximation arises from a saddlepoint approximation to \eikamp, and $x_\perp$ is naturally identified with impact parameter.  This most experimentally accessible regime governs everything from apples to galaxy clusters -- from nonrelativistic processes with ultraplanckian CM energies to, apparently, the case of ultrarelativistic, small angle scattering.

It seems intuitively clear that string effects should not be relevant in the classical regime, and one can understand the broader reason for this with some simple estimates\refs{\LQGST,\GGM}.  First, one could have inserted the string tree-level amplitude into \eikamp, through \eikphase.  
In the high-energy limit this amplitude takes the form
\eqn\fourgrav{T^{\rm string}_{0}(s,t)\propto g_s^2  {\Gamma(-t/8)\over \Gamma(1+t/8)} s^{2+t/4} e^{2-t/4}\ ,}
in units $\alpha'=1/2$.  However, as was explained, the eikonal amplitude is dominated by the saddlepoint given by
\eqn\saddlemom{ q\sim  G_D E^2/x_\perp^{D-3}\ .}
For impact parameters 
\eqn\eikbdy{b\roughly<(GE^2)^{1/(D-4)}\ ,}
the eikonal phase is large at the saddlepoint, $\chi\roughly>1$, indicating that higher terms in the exponential series of \eikamp\ are important.  In particular, we know that the terms with order $N\sim \chi$ give the dominant contribution to the exponential.  Such a term corresponds to a ladder with $N+1$ rungs, and the total momentum transfer is typically expected to divide among these, with a value
\eqn\oneline{ k\sim {q\over N+1} \sim {1\over b} }
per rung.  Specifically, in evaluating the eikonal phase \eikphase\ with the string tree amplitude \fourgrav, the typical momentum transfer is of size \oneline, and so one expects the string corrections to be negligibly small for large impact parameters.

In fact, this discussion extends to consideration of arbitrary short-distance effects.  In the large-impact parameter regime under discussion, they appear to make negligible contribution.  To summarize the picture, as we decrease the impact parameter at a fixed ultraplanckian energy, initially the Born approximation is good, but then breaks down where $\chi$ grows to $\calo(1)$, at $b$ given by \eikbdy.  The eikonal approximation then unitarizes the amplitudes\refs{\GiSr}.  But, in the corresponding diagrams dominating the amplitudes, the momentum transfer $q$, even if large, is {\it fractionated} into small momentum transfers $k$ with size \oneline.  Thus, the scattering is dominated by ``soft" physics, and ``hard" effects are not expected to make important contributions.

In particular, one might have been concerned that divergences and nonrenormalizability are relevant to such a discussion of loop diagrams.  Indeed, in making the eikonal approximation \eikamp, one makes the UV behavior of individual loop amplitudes even worse, as is seen in expanding the exponential in \eikamp\ and using \eikexpress\ -- the individual terms have increasingly bad UV divergences.  However, it seems clear that this is {\it not} a good way to evaluate \eikamp.  This is best illustrated by a simple example\refs{\GSA}, like the integral
\eqn\exampint{I=\int_{\Lambda^{-1}}^1db b^3 e^{ig/b^2}\ ,}
where $g$ is a small  constant and $\Lambda$ is a UV cutoff.  This has a similar short-distance structure to the $D=6$ eikonal amplitude.  Expanding the exponential in powers of $g$ yields arbitrarily high-order powers of $\Lambda$.  Yet, this integral can be explicitly evaluated, and has a finite limit as $\Lambda\rightarrow\infty$ -- illustrating the insensitivity to UV effects.   Expanding and integrating term-by-term is a bad way to evaluate \exampint\ or \eikamp.

These statements can also be tested in supergravity, which provides an explicit regulator making loop amplitudes finite (perhaps to all orders\refs{\SGconj}), and yields explicit expressions for low-loop amplitudes.  In fact, the resulting expressions nicely match\GSA\ terms in the eikonal amplitude \eikamp\ at one and two loops, up to corrections suppressed by powers of $-t/s$, and the cutoff dependence.  Note that this is true despite the extra states -- gravitino, etc. -- present in supergravity.  This illustrates a general feature of such long-distance, ultraplanckian scattering, that of {\it graviton dominance}\refs{\tHoodom,\GSA}.  Specifically, since particles couple to powers of the energy given by their spins, the graviton, as the highest-spin state of the supergravity multiplet, dominates the scattering at high energies.

Momentum fractionation is a mechanism to decouple {\it high-energy} scattering from {\it short-distance} properties of the theory.  For example, even very large momentum transfers only appear to probe the gravitational coupling constant at scales given by \oneline.  This mechanism seems to be a central feature in ``UV-IR" mixing.

 \Ifig{\Fig\Hdiag}{The two-loop ``H-diagram"; straight lines are associated with the scattered high-energy particles.}{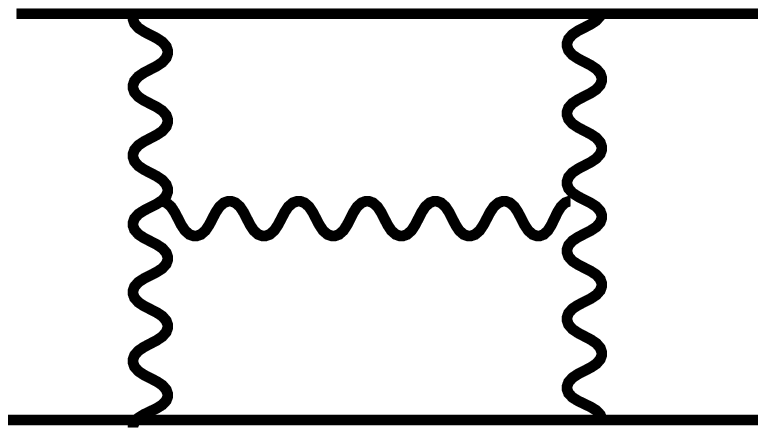}{2}

An important question is when the approximations giving the eikonal amplitude \eikamp\ fail.  One expects both corrections due to radiation, and from higher loops and terms dropped in making the eikonal approximation.  Aspects of the corrections due to radiation are understood, from classical analyses\refs{\EaGi} and study of soft gravitons\refs{\WeinbergNX,\Dono,\GiSr}, and we will focus on the latter corrections.  We have already indicated that they can appear, suppressed by powers of $-t/s$.  These then become important at scattering angle $\theta\sim1$, or, from \saddlemom, impact parameter 
\eqn\schwimp{
b=R(E)\sim (G_D E)^{1/(D-3)}\ ,}
indicated in \Scattdiag.
For example, simple power-counting arguments (see, {\it e.g.}, \refs{\GiPo}) applied to the loop diagram of \Hdiag\ shows that it is suppressed by $[R(E)/b]^{2(D-3)}$ relative to the Born amplitude \born; this also was seen explicitly in the supergravity analysis\GSA.  The presence of such corrections suggest the breakdown of the perturbative expansion at the impact parameter \schwimp.\foot{More precisely, we might think about the perturbative expansion around the classical geometry corresponding to the saddlepoint in the eikonal amplitude.  This will be described below.}
This is our first indication that we lack a means to compute the gravitational S-matrix in this region.

\subsec{The strong gravity regime and the unitarity crisis}

The nature of the perturbative breakdown may be readily understood through comparing a simpler calculation.  We are describing the gravitational field sourced by two particles with high CM energy, but one can similarly examine diagrams sourced by a single {\it massive} particle at rest.  In this case, Duff\refs{\Duff} showed that the sum over tree diagrams gives the expansion of the Schwarzschild metric about flat space in powers of $G_D$.  This expansion manifestly diverges at radii $r\leq R(M)$, where $R(M)$ is the Schwarzschild radius, with parametric dependence as in \schwimp.  A black hole is not a small perturbation about flat space.

As a check of this interpretation, note that we have already indicated the connection of the eikonal amplitudes with the classical approximation.  The classical metric of a high-energy source is well-approximated by the Aichelburg-Sexl (AS) metric\refs{\AiSe},
\eqn\ASmet{ds^{2} = -dx^+ dx^- + dx^{i} dx^{i} +\Phi(\rho) \delta (x^-) dx^{-2}\ ,}
with
\eqn\fd{\eqalign{ \Phi(\rho) &=  -8G_D\mu\ln\rho\quad ,\quad D=4\ ,\cr
&= {16\pi G_D\mu\over \Omega_{D-3} (D-4) \rho^{D-4}}\quad ,\quad D>4\ ,}}
and $x^\mu=(x^+,x^-,x^i)$, $\rho^2=x^ix^i$, and $\mu=E/2$.  
Indeed, scattering of a test particle in this metric straightforwardly yields a scattering angle agreeing with \saddlemom.  Outside the future lightcone of the collision of the two sources, the classical geometry of the two-particle state is that of two such metrics, with centers offset by the impact parameter.  This classical geometry has been shown\refs{\EaGi} (with recent confirmation by numerical relativity\refs{\NumBH}) to form an apparent horizon, thus a black hole, for impact parameters $b\roughly< R(E)$.

So, we expect the sum over tree diagrams connecting the high-energy sources to give the classical metric in this case as well, and that the breakdown of the perturbative expansion is associated with the classical black hole formation.\foot{Note that one can consider either tree diagrams with all but one free ``field-point" leg connected to the high-energy sources, which gives an analog to the calculation of the classical metric of \Duff, or with all legs connected to the external sources, corresponding to contributions to the two-two amplitude like in fig.~2.  These are clearly related, and argued to have corresponding divergences at \schwimp.}

In the case of the eikonal amplitudes, the expansion parameter, there $\chi$, likewise became large, and a more careful way to handle that is to sum up the ladder diagrams to get the classical geometry of the Aichelburg-Sexl collision, and then do a perturbative expansion about this classical geometry.  One might guess that a similar procedure could be followed in the strong-gravity case as well; in particular \refs{\Venezfriends} has suggested that the series can be resummed.  However, comparison with the story of the point mass indicates the apparent outcome: summing the tree diagrams corresponds to building up the classical geometry of the collision and the resulting black hole.  One could then try to perturbatively quantize fluctuations about this geometry.

Here we encounter the unitarity problem, which is so profound it has lead to the apparent paradox of black hole information.  Specifically, such a perturbative quantization was carried out by Hawking, with the conclusion that black holes evaporate\refs{\Hawkrad}, and that the final state is mixed so that information is destroyed and unitary evolution fails\refs{\Hawkunc}.  This analysis can be carried out even more explicitly in two-dimensional models \refs{\CGHS,\GiNe}.

 \Ifig{\Fig\Niceslice}{A Penrose diagram of an evaporating black hole, with one of a family of nice slices pictured.}{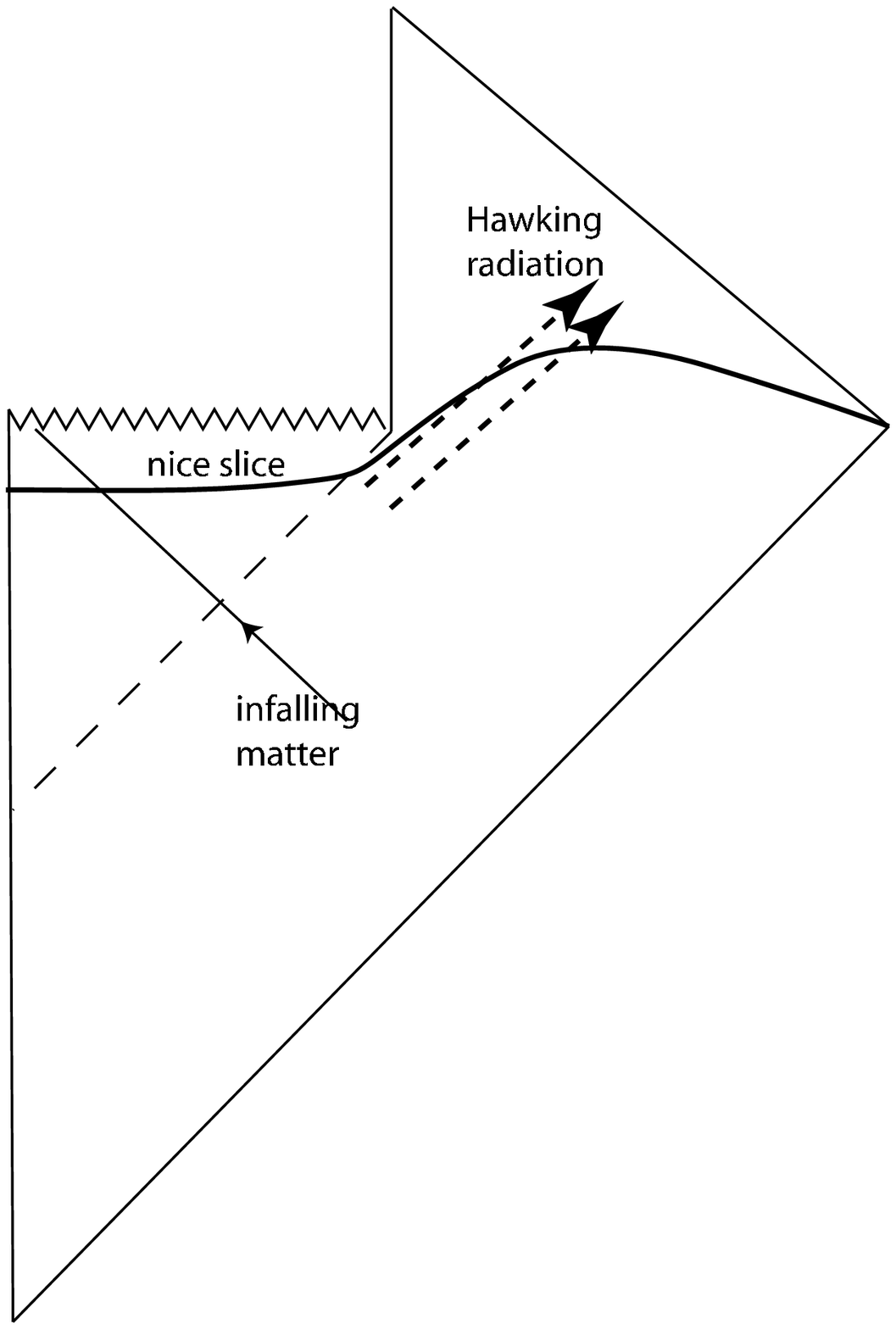}{2.5}

This is not the place for a complete review (see \refs{\Astrorev,\SGinfo}), but the basic arguments are as follows.  First, a family of spatial slices can be drawn through the black hole, avoiding the strong curvature near the classical singularity, and crossing the horizon and asymptoting to a constant-time slice in the asymptotic flat geometry.  One such slice is sketched in \Niceslice. These ``nice slices" were described in \refs{\LPSTU}, with one explicit construction in \refs{\GiNLvC}.  Locality in field theory tells us that the state on such a slice can be represented as a sum of products of states in two Hilbert spaces, one corresponding to inside the black hole, and one outside.  Schematically, this can be written (for more detail, see \GiNe)
\eqn\nicestate{|\psi_{NS}\rangle \sim \sum_i p_i |i\rangle_{in} |i\rangle_{out}\ .}
A description of the state outside the horizon is given by the density matrix formed by tracing out the inside states:
\eqn\rhohr{\rho_{HR}\sim {\rm Tr}_{in} |\psi_{NS}\rangle\langle \psi_{NS}|\ .}
This is manifestly a mixed state.  In fact, one can trace over all degrees of freedom that have not left the vicinity of the black hole by a given time; {\it e.g.} one can describe the density matrix on future null infinity as a function of the corresponding retarded time $x^-$.  This density matrix has an entropy $S(x^-)= -{\rm Tr}[\rho(x^-)\log\rho(x^-)]$, and one straightforwardly estimates that it grows with $x^-$ to a value of order the Bekenstein-Hawking entropy $S(E)\propto G_D[R(E)]^{D-2}$ at a retarded time corresponding to the time the black hole evaporates to close to the Planck size.  This represents a huge missing information, which is also largely independent of what was thrown into the black hole.  The result is a modern update of Hawking's original argument\Hawkunc\ for a unitarity crisis.

The problem with the story is that quantum mechanics is remarkably robust.  Hawking proposed a linear evolution on density matrices, generalizing the S-matrix, but Banks, Peskin, and Susskind\refs{\BPS} argued that this causes severe problems with energy conservation.  A basic argument is that information transfer and loss requires energy loss, and that once allowed, energy non-conservation will pollute all of physics through virtual effects.  In particular, \BPS\ leads to the conclusion that the world would appear to be thermal at a temperature $T\sim M_D$.

The obvious alternative is that information does not escape the black hole until it has reached the Planck size, where the nice-slice argument manifestly fails.  However, with an available decay energy $\sim M_D$ and an information $\Delta I\sim S(E)$ to transmit to restore purity, on very general grounds this must take a very long time, $\sim S^2$.  This implies very long-lived, or perhaps stable, remnants, which come in $\calo(\exp S(E))$ species to encode the large information content.

Since one can in principle consider an arbitrarily large initial black hole, the number of species is unboundedly large.  This leads to unboundedly  large inclusive pair production in generic processes with total available energy $E\gg M_D$, as well as problems with inconsistent renormalization of the Planck mass\refs{\Susstrouble}, {\it etc.}  (The former problem can be seen particularly clearly in the charged black hole sector\refs{\wabhip}, where one considers Schwinger production.)  

So, information apparently cannot get out of a black hole, cannot be lost, and cannot be left behind, and this is the essence of the ``paradox."  It represents a deep conflict between basic principles of Lorentz/diffeomorphism invariance (on a macroscopic scale), quantum mechanics, and locality (also on a macroscopic scale).

Given the apparent robustness of quantum mechanics and of Lorentz invariance, and the puzzling role of locality in gravity, it is natural to propose that the latter is not sharply defined, and that this underlies an explanation of how unitarity is restored.  If this is true, and black hole formation and evaporation is a unitary process, Page\refs{\Page} has argued that information must begin to be emitted by the time scale 
\eqn\pagetime{t_{Page}\sim R(E)S(E)\ ,}
 where the black hole has radiated an $\calo(1)$ fraction of its mass.  This indicates a needed breakdown of the nice-slice argument, and some departure from locality as described in the semiclassical picture of \Niceslice,  over distances comparable to the black hole size, $\sim R(E)$ -- which can be a macroscopic scale. 

While other surprises in high-energy gravitational scattering are not unfathomable -- we are still checking aspects of the picture outlined above -- it seems clear that black holes also form in collapse of massive bodies, yielding a variant of the preceding argument for a conflict between basic principles.

We will return to general questions with, and constraints on, this picture later, but the immediate question is: what does string theory offer to improve it?

\subsec{A role for strings?}

Strings are extended objects, and this suggests that string theory does not have the same locality properties as quantum field theory, and that the preceding picture could be modified.  Indeed, one might expect that strings can become highly excited and extended in a high-energy collision, and it has occurred to various string theorists that this could change our story.  In particular, it was suggested in \LPSTU\ that string excitation and extendedness could play a role in breakdown of the nice-slice argument, and in \refs{\Venezesc} that it could play a role in how information escapes a black hole.

We have already given an argument that string corrections to the tree-level amplitude do not significantly change the eikonal amplitude.  However, another effect that we have neglected is excitation of a scattered string through coherent interaction with the gravitational field of the other high-energy string.  Such effects were found to be present in \refs{\ACV}.  They were later interpreted as due to tidal excitation\refs{\LQGST}:  if a composite object, whether a hydrogen atom or a string, scatters off the AS metric \ASmet, it experiences tidal forces, and can become excited.  In the string case, there is a critical impact parameter, indicated by the dotted line in the right half of \Scattdiag, and given by\refs{\ACV}
\eqn\critimpact{b_t\approx {1\over M_D} \left({\sqrt{\alpha'}E}\right)^{2/(D-2)}\ ,}
where this becomes an important effect.  The most extreme interpretation of this observation is that, due to string excitation, the picture we have given of gravitational scattering cannot be trusted, and is dominated by excited string effects.

This can be tested by explicit study of string scattering in an AS metric.  Specifically, one can adopt a ``probe" approximation, in which the scattered string's backreaction on the metric is neglected.  Such an approximation naturally emerges at leading order in an expansion in small scattering angle $\theta$, and can be used to deduce the basic parameters governing excitation.  The problem was studied in \refs{\GGM}, which took advantage of the natural light-cone structure of the AS metric \ASmet\ to explicitly quantize the string.  The picture of tidal excitation was confirmed, as was the intuitive picture\refs{\LQGST} that, once the string receives a tidal ``kick" from the metric, it spreads out only on a characteristic timescale, computed in\GGM:
\eqn\stringspread{t_t\sim {b^{D-2}\over G_D\sqrt{\alpha'}E^2}  \ .}
In short, the string appears to behave in accord with basic intuitions of causality, and is not instantaneously spread out by the collision. 

This can be compared with the black hole formation.  In a classical collision, in fact the black hole (apparent horizon) forms {\it before} the AS shock waves meet, by a time of order $R(E)$.  Combining these observations, one has a picture in nice accord with causality.  If two high-energy strings collide at impact parameters between $b_t$ and $\sim R(E)$, they become excited, and this is important in description of the asymptotic state.  (With finite string coupling, one expects decay to multiple asymptotic strings.)  But, for impact parameters $\roughly< R(E)$, a black hole forms {\it before} the strings meet the gravitational shock wave of their counterpart.  When they do meet, the strings becomes excited.  But, they are then inside a black hole.  So far string behavior has been found to respect basic intuitions of causality, and in particular, we don't have any good evidence an excited string can escape a black hole -- it would appear to behave a lot like any other extended object.

While these statements do not rely on complete calculations, they do extrapolate sharp calculations, and there is no sharp picture of effects that would modify them.  In short, there is no clear mechanism to prevent black hole formation, or to permit information escape, based on the extendedness of strings.  

So far, we appear to conclude that string theory in fact does not lead to major modifications of the picture outlined in the preceding sections, and summarized in \Scattdiag, aside from the tidal excitation of asymptotic states in the  regime indicated.  In fact, this story is related to the momentum fractionation described above, limiting the relevance of short distance effects. 

Moreover, string theory appears to encounter the same divergence in the perturbation series that was described above, at $b\sim R(E)$, associated with black hole formation -- we have found no way to avoid it.

\subsec{Summary}

To summarize the situation, ultraplanckian scattering in the strong-gravity regime manifests a unitarity crisis.  This appears to be associated with a breakdown of perturbation theory.  These are long-distance issues, and appear essentially independent of the question of perturbative renormalizability -- which has been a big motivator for string theory and other approaches to quantum gravity -- or of other short distance effects.  (For the same reason, one does not expect a rescue from loop quantum gravity, or asymptotic safety\GSA.)  Specifically, there are no indications that even order-by-order finiteness in perturbation theory, as is expected in string theory (and possibly in 
 in $N=8$ supergravity\refs{\SGconj}) addresses the issue.  We face what is apparently a non-perturbative problem, which seems to be a centrally profound aspect of the problem of reconciling quantum mechanics with gravity.

\newsec{Is there a decodable hologram?}

If the improved behavior of perturbation theory in string theory doesn't address the issue of computing the S-matrix in the strong gravity regime, we apparently need a fully non-perturbative definition of the theory.  It has been proposed that dualities, most notably AdS/CFT, furnish such a definition.  Indeed, for purposes of defining the S-matrix this seems  the simplest duality, and other approaches appear to offer more difficulties, so it will be examined as a proxy for more general dualities.

AdS/CFT has been suggested as a concrete implementation of the proposal of {\it holography}\refs{\thooholo,\sussholo}, which in its strongest form states that a gravitational theory governing a region of space has a completely equivalent theory on the lower-dimensional boundary of that region.  This proposal  gives an idea for a kind of nonlocality that might address the unitarity problem of black holes. And, in AdS/CFT the boundary theory is ${\cal N}=4$ super Yang-Mills with group SU(N), a theory that is well-defined in the sense that we could presumably put it on a computer, and so if the holographic proposal holds, this would give a definition of string theory in asymptotically AdS space.

Clearly one would like to better understand the precise relationship between the theories, and this has been the subject of a lot of exploration.  Our specific question is in what sense there is an {\it equivalence} between the theories, and whether this defines a good approximation to the flat-space S-matrix.  If it does, we should be able to examine the properties of this S-matrix, and understand the underlying non-local physics avoiding the unitarity problem.  This is part of the problem of ``decoding the hologram."  A possible alternative to consider\refs{\FSS} is that AdS/CFT is more like a projection, in the sense that not all of the information of the bulk theory is contained in the boundary theory; for example, the latter might only capture an appropriately coarse-grained description of the bulk physics, perhaps somewhat like low-energy effective field theories only capture the IR dynamics, and ``forget" details of the short-distance physics.

If there is an equivalence, the boundary theory has a Hilbert space which we would expect to correspond to that of the bulk, via a map
\eqn\hilbmap{ M: \calh_B\rightarrow \calh_\partial}
that is one to one and unitary.
The boundary theory also has a hamiltonian and thus unitary evolution operator $U$, which we would like to describe corresponding objects in the bulk.  In particular, such a map $M$ 
can be used to define a bulk evolution operator $U_B=M^\dagger UM$ which is unitary, $U_B^\dagger U_B=1$.

The correlators in the boundary theory define a sort of AdS analog of the S-matrix\refs{\BSM}.  In a complete equivalence between the two theories, we would also expect to be able to extract a good approximation to the flat space S-matrix, if the AdS radius 
$R=(g_{YM}^2N)^{1/4}l_{string}$ is large as compared to the string length $l_{string}$.  Thus in the large-$R$ limit we would expect $U_B$ to determine $S$, such that $S^\dagger S=1$.

There is no question that there are deep and useful relations between the bulk and boundary theories.  But, the essential question is whether the bulk theory, and in particular its S-matrix, can be reconstructed in detail from the boundary theory.  This has been actively investigated.  The rest of this section will summarize some approaches to this problem, some of the constraints and difficulties encountered, and possible conclusions.

\subsec{Normalizable states}

Considering radial quantization in the conformal (boundary) theory, one can create a state from the vacuum by acting with an operator at the origin of ${\bbb R}^4$, $|\psi_\calo\rangle= \calo(0)|0\rangle$.  This can be conformally mapped to the cylinder, ${\bbb R}\times S^3$, with the operator acting at $t=-\infty$.  Under \hilbmap, such states should correspond to states of theory in the bulk of AdS, with ${\bbb R}\times S^3$ identified as the boundary.  

In the case where the bulk string coupling, $g=g_{YM}^2$, is taken to be zero, there have been checks that the resulting spectrum of free-particle states indeed corresponds to states created by such boundary operators; for a review, see \MAGOO.  However, given the relative triviality of the free theory, a large part of this statement seems kinematical.

 \Ifig{\Fig\adscoll}{AdS behaves like a gravitational ``box" of radius $R$; normalizable states correspond to particles confined to the box, which undergo multiple collisions.}{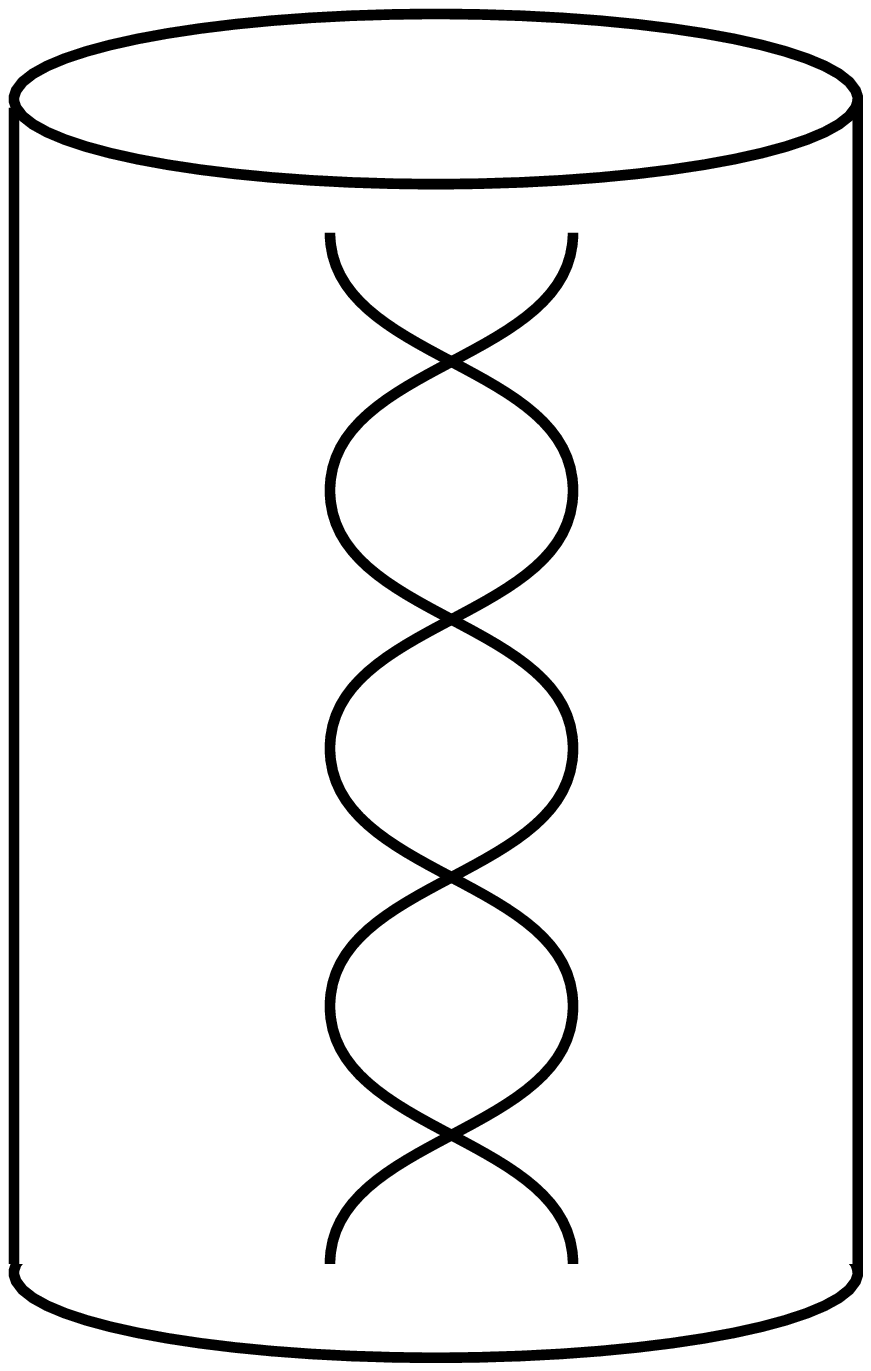}{2}
 
To isolate the S-matrix, one needs to consider the interacting theory, say with finite but small $g$; one naturally interprets \hilbmap\ as a map between the states of the boundary theory and the {\it interacting} states of the bulk theory.  However, this adds an essential complication.  In flat-space scattering, one can arrange a two-particle state such that the particles are well-separated in the past, then collide, and scatter into other asymptotically separated particles.  But, for normalizable states in AdS, one expects trouble with such asymptotic separation.  Namely, as pictured in \adscoll, such normalizable states are effectively confined to a box of size $R$ for infinite time, and hence undergo infinitely many collisions.  In this sense, interactions are not necessarily small -- even for $g\ll 1$ -- for such states.

If we imagine that the AdS radius is large, say $10^{10}$ light years, what we would like to do is isolate the effects of one such collision, which might then be interpreted as giving a controlled approximation to the S-matrix.  This suggests that one needs an analog of the LSZ formalism of flat-space theory, in order to do so.  Specifically, we would like to find a state of the fully interacting Hilbert space $\calh_B$ that, at $t=-1$ year, looks like two incoming, essentially free particles, say in specific wavepackets, and at $t=+1$ year looks like a state of some outgoing, far-separated particles.  

If the state is normalizable and obtained from the preceding operator-state correspondence, this is challenging.  The problem is unscrambling the infinite interactions to the past (or future) of our collision region.  Specifically, if we attempt to use the $g=0$ correspondence to tell us how to construct initial two-particle wavepackets of the desired kind, the corresponding state in the interacting theory will receive corrections from the infinite reflections and interactions pictured in \adscoll.  

The problem of isolating states of the kind described, in order to approximate the S-matrix, remains an open question.  One recent proposal\refs{\Katzetal} suggests a way scattering amplitudes (reduced transition matrix elements) might be extracted, but it is not clear that it truly overcomes the ``multiple collision" problem, and suffices to fully determine a close approximation to the S-matrix.

\subsec{Boundary compact states}

An apparently more promising proposal for defining the S-matrix was made by Polchinski\refs{\PolS} and Susskind\refs{\SussS}: excitations created at the boundary of AdS can be allowed to scatter, and then the S-matrix can be ``measured" by boundary operators sensitive to the state after the collision.   Specifically, if $b$ denotes a boundary point, one can integrate against a source function $f(b)$,
\eqn\bdysource{\int db f(b) \calo(b) |0\rangle}
to create such an excitation.

Suppose we choose coordinates where the $AdS_{d+1}$ metric is
\eqn\adsmet{ds^2={R^2\over \cos^2\rho}\left(-d\tau^2 + d\rho^2 + \sin^2 \rho d\Omega_{d-1}^2\right)\ .}
The prescription found by Gubser, Polyakov, Klebanov, and Witten\refs{\GKP,\WittAdS} then
gives the bulk wavefunction of the field corresponding to $\calo$, via the bulk-boundary propagator:
\eqn\bdywavepack{\psi_f(x) =\int db' f(b') G_{B\partial}(b',x)\ . }
The asymptotic behavior of the  wavefunction corresponding to \bdysource, with operator dimension $\Delta$ (related to the bulk mass by $2\Delta = d+\sqrt{d^2 +4m^2R^2}$), is
\eqn\psinn{\psi_f(x)\limrho {(\cos\rho)^{d-\Delta}\over R^{(d-1)/2}} f(b)\ .}
This  has nonnormalizable behavior when approaching points on the boundary in the support of $f$, and so in order to create a state of the bulk Hilbert space, in the sense that it is normalizable at some time $\tau$, $f$ must have compact support.

 \Ifig{\Fig\AdSS}{The picture for a proposed derivation of the S-matrix from boundary-compact sources which produce particles at spatial infinity.}{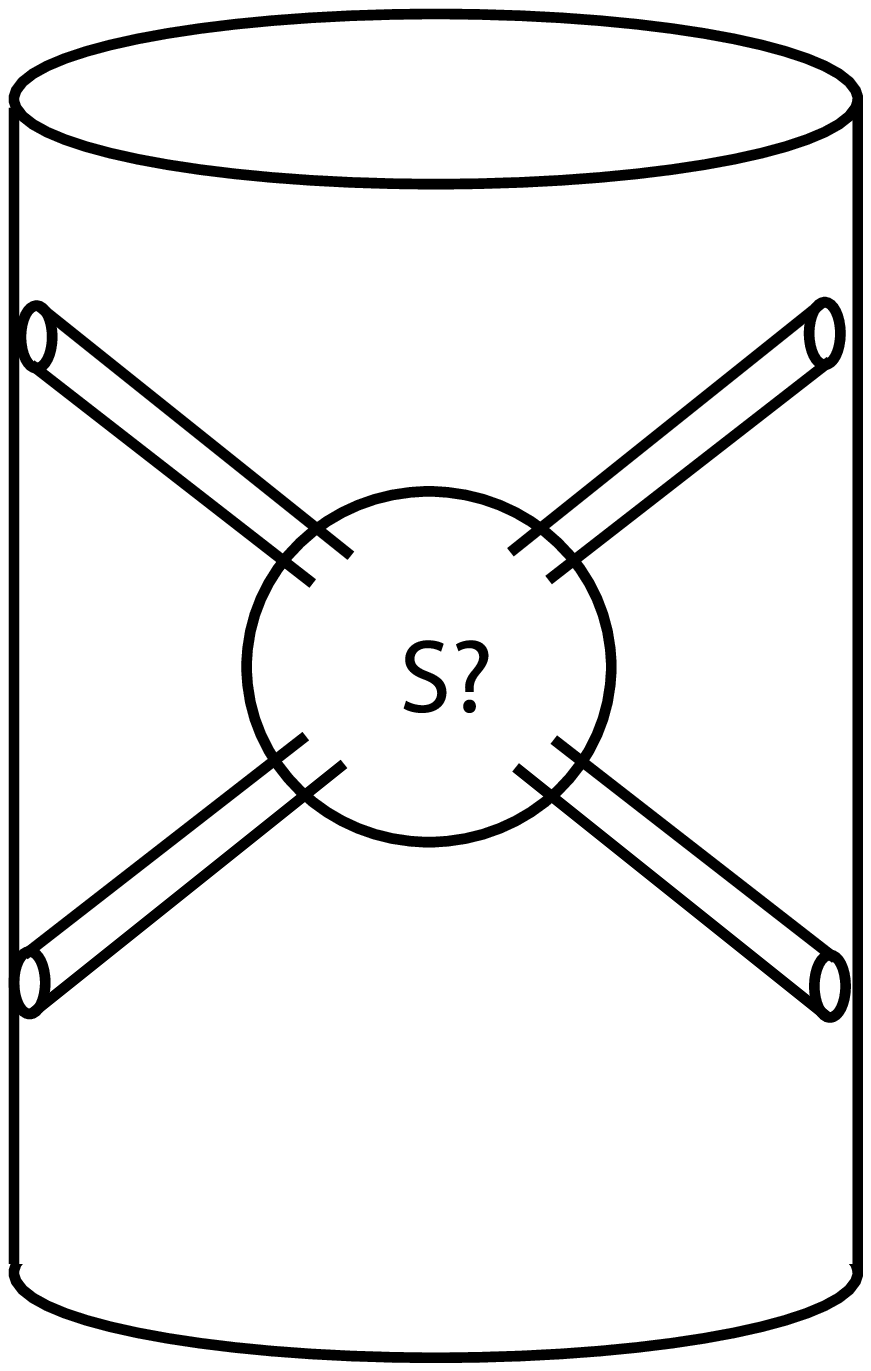}{2}

Ref.~\refs{\GGP} gave a construction of such sources designed to produce localized wavepackets in the bulk, and to test the proposal to extract the S-matrix as sketched in \AdSS.  Specifically, these ``boundary-compact" sources take the form
\eqn\bcsource{f(b) = L(\tau, \ehat) e^{\pm i \omega R \tau}\ ,}
where $L(\tau, \ehat)$ is sharply peaked near a point $b_0=(\tau_0,\ehat_0)$, with widths $\Delta \tau$ and $\Delta \theta$, and has compact support, and $\omega$ gives the typical energy of the wavepacket seen by a central observer at $\rho\ll 1$.

For such narrowly-peaked sources (compact support or not), a simple, explicit expression can be found for the corresponding wavepackets in the flat region $\rho\ll1$, generalizing \refs{\GaGi}.    
This follows from a Schwinger-like expression for the bulk-boundary propagator,
\eqn\gbdelr{G_{B\partial}(b',x) = { (\cos\rho)^{\Delta}{\hat N} \over R^{(d-1)/2} }\int_0^\infty d\alpha \alpha^{\Delta-1} \exp\left\{i\alpha [\cos(|\tau-\tau'|-i\epsilon) - \sin\rho\, \ehat\cdot\ehat']\right\}\ ,}
where 
\eqn\nhatdef{{\hat N} = {2\Delta-d\over i^{\Delta-1} 2^{\Delta+1} \pi^{d/2} \Gamma(\Delta+1-d/2)}\ .}
Specifically, using \bdywavepack, expanding the exponent of \gbdelr\ about $\tau'=\tau_0$, setting $\alpha = kR$, and defining
\eqn\barcoords{t = -R \cos(\tau -\tau_0)\ ,\ r = R \sin\rho}
gives a bulk wavepacket
\eqn\flatpack{\psi_f = \int_0^\infty {k^{\Delta-1}dk\over 2k}\int d\ehat'\ F(k,\ehat', t,r) e^{-ikt-ik\ehat'\cdot {\vec r}}\ ,}
with
\eqn\Fdef{F(k,\ehat',t,r) = {2k{\hat N}\left[1-r^2/R^2\right]^{\Delta/2} \over R^{(d-1)/2-\Delta}} \int d\tau e^{ik\sqrt{R^2-t^2}(\tau-\tau_0+i\epsilon)} f(\tau,\ehat')\left\{1+\calo[kt(\tau-\tau_0)^2]\right\}\ .}
In the coordinates $r,t$, the AdS metric
\eqn\rtmet{ds^2 = {1\over 1-r^2/R^2} \left(-{dt^2\over 1-t^2/R^2}+ {dr^2\over 1-r^2/R^2} + r^2 d\Omega^2_{d-1}\right) }
is a good approximation to the Minkowski metric for $t,r\ll R$.  In this region, and for $kt\ll 1/(\Delta\tau)^2$, $F$ in \Fdef\ is basically just the Fourier transform of the source $f$ with respect to time.  And,  for $m=0$, thus $\Delta=d$, \flatpack\ gives the standard Fourier representation of the bulk wavepacket.

In order to describe bulk states, $f$ is required to have compact support.  Then $F$ will not have compact support.  This excludes the kinds of wavepackets used in careful treatments of scattering\ReSi, specifically the {\it regular} wavepackets, which have support that is compact in momentum space and excludes zero frequency.  Conversely, compactly supported $F$ means noncompact $f$.\foot{One also might try semicompact sources like those described in \MaroUH.  The expression \Fdef\ would not give compact $F$, but needs to be modified to take into account the extent of the source.  This remains an open question, though one also expects to encounter difficulties either like those outlined in the previous or next subsections.}

This appears to be an obstacle to approximating the usual Hilbert space of bulk states, in the limit of large $R$.  One can ask how boundary-compact wavepackets are different from more familiar ({\it e.g.} gaussian) wavepackets.  They do have characteristic longitudinal and transverse widths, given by
\eqn\widths{\Delta t\approx R\Delta \tau\ ,\ \Delta x_\perp \sim 1/\Delta p_\perp\sim 1/(\omega \Delta\theta)\ .}
But outside these widths, one finds\refs{\GaGi} that they have power-law tails, with characteristic falloff in both longitudinal and transverse directions
\eqn\psiasymp{\psi_f \sim 1/r^\Delta\ .}

Without the ``sharper" regular wavepackets, one should reconsider how to give a careful description of scattering; there is potential for trouble in precisely approximating the S-matrix.
For example\GaGi, these tails hinder separation of the ``scattered" contribution from the ``direct" contribution to S.  This issue might be addressed by subtraction of the disconnected correlators directly in the boundary theory,\foot{I thank T. Okuda and J. Polchinski for discussions on this point.} {\it e.g.} for the four-point function
\eqn\direct{\langle \calo_1\calo_2\calo_3\calo_4\rangle_c = \langle \calo_1\calo_2\calo_3\calo_4\rangle - \langle \calo_1 \calo_2\rangle\langle \calo_3 \calo_4\rangle - \langle \calo_1 \calo_3\rangle\langle \calo_2 \calo_4\rangle-\langle \calo_1 \calo_4\rangle\langle \calo_2 \calo_3\rangle\ .}

But, returning to one of our basic motivations, note that one expects the gravitational S-matrix in the strong gravity regime to have entries {\it exponentially small}  in the entropy, which is a power of the energy.  Specifically, the amplitude to form a black hole and then have it decay back to a typical multi-particle final state is of order $e^{-S/2}$.  The ``meat" of the unitarity problem is in matrix elements of this size;  making errors in the density matrix of the final state of this size can make the difference between having a pure state and a mixed state like described by Hawking.  With a Hilbert space including gaussian wavepackets, one can approach this kind of resolution.   However, the tails \psiasymp\ would appear to impede resolution of such small S-matrix elements.  Specifically, they are expected to give larger amplitudes for two colliding particles to {\it miss} entering the strong gravity regime, and instead scatter by a longer-range gravitational process.\foot{For a gaussian wavepacket, the longitudinal width of the packet must satisfy $\Delta t \roughly> 1/\Delta E > 1/E$, which places a bound on the amplitude for tails of wavepackets colliding at zero impact parameter to be ``outside" the horizon:  ${\cal A}\sim \exp\{-[R(E)]^2/\Delta t^2\}\roughly > \exp\{-S^2\}$.   Similar arguments can be given for transverse widths.}

Even if one puts these questions aside, a puzzling question is how interactions purely in the boundary theory could ``mock-up" the correct local scattering in the bulk.  For example, if by adjusting $f$ and thus $F$, we change the trajectory of a particle slightly, $\delta ({\rm impact\ parameter})<<< R$, so that it no longer collides with another, the scattering behavior is completely different.  Yet, in the boundary theory, the corresponding state is spread in a complicated way over the boundary at the collision time\GaGi.  

Some more precise necessary conditions on the requisite boundary behavior can be found.  For example, in order to correctly reproduce a momentum-conserving delta function, in a plane-wave limit, the four-point correlator \direct\ should have a specific kind of singular behavior described in \GGP.  This is not directly related to the operator-product limit where two boundary points coincide, but occurs when the boundary points are light-like related to a common bulk point.  Moreover, to produce {\it e.g.} the Born behavior \born, the correlator must also have a specific kind of subleading singularity.   If these behaviors are present, one can extract the reduced transition matrix elements, in a plane-wave limit\GGP.
One can ask about the origin of such singular behavior, from the boundary perspective.  In particular, \refs{\HPPS} has formulated a conjecture that a CFT that has a large-N expansion, and in which single-trace operators of spin $>2$ have parametrically large dimensions, will exhibit the leading singularity, and has given some supporting evidence for this conjecture.  But, in view of the above discussion, the question of extracting the full S-matrix this way remains an open problem.

\subsec{Noncompact sources}

The problematic nature of boundary-compact wavepackets encourages closer examination of sources without compact support.  From the boundary perspective, these do not create states, but rather deformations of the theory.  In particular, quantities like 
\eqn\diverg{\langle \calo(b)\int db' f(b') \calo(b')\rangle\ ,}
diverge.
These also do not create states in the original bulk Hilbert space;  total particle number diverges near the boundary.  This latter property naively impedes focussing of interactions near the ``center" of AdS, as we needed to derive the S-matrix. Specifically, bulk correlators will include expressions like
\eqn\divexp{\int dV \, \psi_{NN} \psi_{NN} G_{Bulk}\ ,}
where the bulk Green function is integrated against a pair of non-normalizable wavefunctions, producing a divergence near the boundary, corresponding to infinite interaction amplitude there.  

However, gravity is special; its effective coupling  is proportional to the {\it energy}.  Specifically, the (super)graviton couples through the (super)stress tensor.  For a state with a fixed energy  at the center of AdS, the proper stress tensor seen by observers near the boundary redshifts to zero as the boundary is approached, allowing the analogous expression to \divexp, with stress tensor integrated against graviton propagator, to converge\refs{\GaGitoapp}\foot{I thank J. Polchinski and M. Gary for discussions on this point.}  despite the infinite particle number.  This allows finite integrals of the corresponding connected four-point function \direct\ against non-compact sources.

An interesting question is how such quantities could be interpreted as a bulk S-matrix, in view of the nonnormalizability.  One apparently lacks a normalization condition, and more generally formulating the unitarity condition $S^\dagger S=1$, which apparently requires a sum over intermediate normalizable states, is a challenge.  These issues (and a related question of whether an analogous unitarity condition can be formulated directly for correlators) will be discussed elsewhere\GaGitoapp.

\subsec{Summary}

I have summarized investigation of the problem of ``decoding the hologram," specifically by extracting the bulk S-matrix.  There are a variety of necessary conditions that need to be satisfied.  In particular, one would need the boundary theory to not only have a large-N expansion, but also to yield a specific hierarchy of dimensions and produce certain specific leading and subleading singularities\refs{\GGP,\HPPS}.  The boundary dynamics needs to reproduce interactions that behave approximately locally from the bulk viewpoint.
And, one seemingly needs a method to isolate good approximations to the flat S-matrix.  This requires a resolution of either the ``multiple-interaction problem," the ``tail problem," or the ``non-normalizability problem" of the preceding sections.   

Unitarity of the boundary theory is often heard equated to unitarity of the bulk theory, but as has been illustrated, attempts to implement such statements in detail to approximate a unitary bulk S-matrix run into difficulties.  One may view these as merely technical difficulties -- in which case one should resolve them.  

Or, perhaps they are of a more fundamental nature, and are telling us that the AdS/CFT correspondence does not define all the quantities we need to describe the  bulk theory.  Clearly the correspondence has a lot of power, and has produced surprising connections between bulk and boundary physics.  However, these may (at least partially) result from universality and symmetry, which are powerful principles.  It is not clear {\it why} the boundary theory should completely produce bulk physics with the correct (approximately) local interactions.  An alternative possibility\FSS\ is that the boundary theory provides a sort of effective theory, or thermodynamic description, which only furnishes a kind of coarse-grained version of the bulk dynamics, and in particular does not give a fine-grained S-matrix satisfying $S^\dagger S=1$.  We do not yet have a ``no-go" theorem for decoding the hologram -- but we have found some apparently significant challenges to the proposed equivalence.

Indeed, if the correspondence were complete, we would expect it to address a kind of question on which it has so far been largely silent.  Namely, in physics we make observations that are approximately local in spacetime, and we do not observe an S-matrix at infinity.  We turn next to this question.

\newsec{The problem of local observables}

To explain what we see, physics apparently needs a way of describing the observations of a localized observer, whether that observer is viewing fluctuations in their expanding cosmos, or falling towards the center of a black hole.  While there are some ideas how one might find certain  localized observables in AdS/CFT (for example, associated with positions of branes), this is very much an unresolved problem.

Indeed, this is a general problem in gravity.  While in field theory one can construct local gauge-invariant observables from fields, the low-energy gauge symmetry of gravity is diffeomorphisms, under which such a local operator will not be gauge invariant:
\eqn\diffop{\delta_\xi O(x) = \xi^\mu\partial_\mu O(x)\neq0\ .}
String theory largely sidesteps this question, in describing an S-matrix.

Since we seem to need to make predictions about such observations within an ultimate theory containing gravity, particularly in the cosmological context, we expect a complete theory of quantum gravity to explain how such local observables arise, at least in an approximation.  Note that we also expect this from correspondence of such a theory with field theory in the correspondence limit  $G_D\rightarrow0$.  

The idea for how this  could work goes back to at least Leibniz and Einstein, with modern additions by DeWitt and others.  Specifically, localization of phenomena in space and time is taken to be in {\it relation} to other features of the state describing the Universe.  We can explain some of the ideas for how this works, in the context of a low-energy effective description, although a key question is how these ideas are implemented in a more fundamental and complete theory.  In particular, the diffeomorphism symmetry implies that configurations must satisfy the constraint equations of general relativity, which in the quantum context give us the Wheeler-DeWitt equation.  We can have a solution to this equation that nonetheless has non-trivial features in it, {\it e.g.} cosmological fluctuations, galaxies, and planets.  This is somewhat like a spontaneous breaking of diffeomorphism symmetry.  And, in such a state we can localize observables relative to those features.  

Specifically, let $|\Psi\rangle$ denote a solution of the Wheeler-DeWitt equation, 
\eqn\wdeqn{\calh_{WD}|\Psi\rangle = 0\ .}
One seeks operators $\calo$ that are gauge invariant, in particular satisfy $[\calh_{WD},\calo]=0$, and that in appropriate states approximately reduce to local operators,
\eqn\localapprox{\langle\Psi| \calo|\Psi\rangle\approx \langle O(x_0)\rangle\ ,}
where the point $x_0$ is selected by a feature in the state.  Ref.~\GMH\ outlines how such operators might be constructed in the effective field theory approximation, for example as integrals of the form
\eqn\opconst{\calo = \int d^D x \sqrt{-g} B(x) O(x)}
where $B(x)$ is an operator selecting a certain feature of ``background" fields at, say, point $x_0$.  Ref.~\GMH\ contains more detailed examples of these and other kinds of relational observables.  Another example is provided by current studies of cosmological fluctuations, in the semiclassical limit.  The underlying dynamics respects diffeomorphism symmetry, implying in particular a question of how to specify ``time."  In the context of slow-roll inflation, a common choice is to define time slices by constant values of the inflaton field.  This in particular plays a physical role in computing a fluctuation spectrum at ``reheating," which is where the inflaton reaches a specific value $\phi_{RH}$ where inflation ends.

One would like to understand how such ideas could be implemented in string theory, whether in AdS/CFT or in some more general formulation.  If AdS/CFT were to encode all details of bulk dynamics, it might for example be natural to look at certain relationally-specified quantities involving the matrix degrees of freedom  (which are acted on only by the boundary gauge transformations).  
In the absence of such constructions in string theory, it is difficult to view string theory as providing a complete quantum theory of gravity.  

In particular, descriptions of local observations would seem to be needed for a picture that fully addresses the black hole information ``paradox."  The essential tension that we have outlined arises when one attempts to give a description of  both the physics seen by infalling observers, and by observers outside the black hole.  While an S-matrix description of the process of formation and evaporation of a black hole would give very important clues, more is needed in order to understand the precise way in which a unitary picture emerges and departs from the semiclassical (and local) arguments for information loss.  The most promising approach to describing such localized observations appears to be via a realization of the relational ideas that have been outlined.

It is amusing to note in this connection that once again string theory certainly serves as a source of inspiration and ideas:  in particular, since world-sheet dynamics involves an implementation of $1+1$-dimensional diffeomorphism symmetry, we should be able to find such constructions for ``observers" on the string world sheet!  Such a toy model for the basic picture we have outlined was in fact developed in \GaGiobs, based on string methods.  In particular, familiar vertex operators of string theory supply operators of the general form \opconst.

Finally, note that a proper formulation of such operators in the cosmological context might go a long ways towards addressing the ``measure" problems in inflationary cosmology, whose puzzles and paradoxes appear to stem dominantly from large IR effects.  One expects these could be tamed in a consistent and complete formulation of such observables.  

Studying the problem of observables and studying scattering are complementary approaches to the problems of quantum gravity; a better understanding of local observables is likely to shed further light on properties, particularly unitarity, of the S-matrix, and on the other hand understanding scattering is likely to help us with the problems of cosmology.

\newsec{Outlook}

To summarize the situation, string theory has been a continuous source of new ideas in mathematics and physics, and showed a lot of initial promise for resolving the problems of quantum gravity.  However, the more profound problems are yet to be convincingly addressed, and there are deep puzzles about {\it how} they might be addressed by string theory.  

If string theory is to be (at least) a prescription for defining a gravitational S-matrix, it apparently needs to do so in the ultraplanckian regime.  While its addressing nonrenormalizability seemed initially very promising, we have argued that a deeper problem that is exhibited in this regime is that of unitarity.  The corresponding strong gravity dynamics appears to represent a serious breakdown of perturbation theory -- presenting a deep conflict between very basic fundamental principles, and requiring a non-perturbative definition of the theory.

AdS/CFT and similar dualities  were initially hailed as promising ways to provide such a non-perturbative definition.  But, as I have outlined, the problem of extracting detailed bulk quantities -- such as the flat-space S-matrix, reveals significant difficulties.  It remains to be seen if these are surmountable, or whether such dualities only hold at an effective or appropriately coarse-grained level.

In particular, for cosmology and to match experience, we also seek a definition of other quantities that are more general than the S-matrix, and describe local observation, in some approximation.  So far, this question has been largely avoided in string theory.

It is important to consider more generally the outlines of physics that could address these questions, whether this physics is string theory or something else.  One guide is the shape of our ignorance -- where do our familiar constructs of local quantum field theory fail?  Exploring a possible analogy between this and the description of the correspondence boundary between classical physics and quantum mechanics, an answer 
to this question would be an analog of stating the uncertainty principle.  There have been string-motivated proposals of this nature, particularly the string uncertainty principle\refs{\Vene,\Gross},
\eqn\stringup{\Delta X \geq {1\over \Delta p} + \alpha' \Delta p\ .}
But, a different way to look at the problem is to ask when familiar constructs of field theory break down due to strong gravitational effects.  Lorentz invariance suggests that one can boost a single particle to arbitrary energy (and appears at odds with \stringup).  But, if one considers a two-particle Fock space state, for example with particles with approximately definite momenta and positions,
\eqn\twopart{\phi_{k,x} \phi_{k',x'}|0\rangle}
a local quantum-field theoretic description would appear to fail when 
\eqn\Locbd{ |x-x'|^{D-3}> G |k+k'|}
is violated,
where $G\sim G_D \hbar$.  This proposed limitation on the dynamical description of QFT, Fock space states,  is called the locality bound\refs{\locbdrefs}, and generalizes to a statement for $N$ particles\refs{\LQGST}, and for de Sitter space\refs{\GiMa}.  

Merely knowing where local QFT fails is an important clue, but one needs to find the underlying mathematical framework and dynamics of a more complete theory.  Using again the analogy of quantum mechanics, one needed to deduce that phase space is replaced by the Hilbert space of quantum wavefunctions, governed by ({\it e.g.}) Schr\"odinger mechanics.  Another important hint could come from asking how this dynamics could depart from the semiclassical/local QFT description of dynamical geometry.  One does find indications\refs{\QBHB} of a breakdown of this description of black-hole evaporation, on the time scale \pagetime, and of a similar breakdown in inflationary evolution\refs{\QBHB,\GiSl}. 

 In the black hole case, one would like to understand what kind of modification of the semiclassical picture could give rise to a unitary S-matrix.  In particular, there is the picture described as {\it black hole complementarity}\refs{\STU}, stating that observables inside and outside the black hole are not independent, and
this picture fits with the ideas of holography.  But, the picture suggests return of the information carried by  a qubit that falls into the black hole on a characteristic retention timescale $t_{ret}\sim R\log R$ (see, {\it e.g.}, \SekinoHE), much shorter than \pagetime, and thus represents a very radical departure from  Hawking's original prediction based on local QFT, $t_{ret}=\infty$.  A seemingly more conservative possibility\refs{\GiNLvC} is  that the information only leaks out on the timescale given by \pagetime.  
 
Other possibly important clues come from ``parameterizing our ignorance" of an S-matrix governing gravitational scattering, for example in two-two processes\refs{\GiSr,\GiPo}.  If there is such a quantum-mechanical description, and it has basic coarse-grained properties of long-distance gravity, particularly black hole formation and evaporation, this has potentially important implications in properties of the S-matrix and of the underlying theory.   Such a parameterization does suggest nonlocality, making its appearance in the S-matrix through nonpolynomiality.  
In gravity, the important property of crossing symmetry is more subtle than in massive theories, but there are indications of a story consistent with crossing, which then constrains possible acausal behavior.  Put together, one has a suggested   picture where nonlocality is allowed without encountering acausality\refs{\GiPo,\GiNLvC,\Erice}.

A clear question is to provide a sensible mechanical description of such physics, which would stand at odds with the usual local picture when viewed from the perspective of the semiclassical geometry of the evaporating black hole.  As we have described above, locality is not expected to be a precise notion in quantum gravity; in particular, if localization is only defined relative to other features in the state, those features can fluctuate as well.  But, tampering with locality is dangerous -- in field theory, locality and causality are intimately linked, and acausality appears to imply inconsistency.   Nonetheless, as indicated, it may be possible to walk a fine line between locality and inconsistency.  

We seek a consistent framework for describing quantum processes, in which spacetime locality emerges in an approximation.  This, together with the requirement that it produce an S-matrix (and other local dynamics) with familiar properties of gravity seems a very tall order.  This is actually encouraging, as it suggests the problem is sufficiently constrained to guide the resolution of this profoundly challenging set of problems.  It remains to be seen what role string theory plays in this, and whether it can provide further clues.

\bigskip\bigskip\centerline{{\bf Acknowledgments}}\nobreak

This work  was supported in part by the Department of Energy under Contract DE-FG02-91ER40618.  I thank G. Horowitz and D. Marolf for comments on a draft of part of this paper, and M. Gary and J. Polchinski for useful discussions.

\listrefs
\end